# Nuclear matter for compact stars and its properties

**Partha Roy Chowdhury[1]*†**

*Department of Physics, University of Calcutta, 92, A.P.C. Road, Kolkata-700 009, India*
†*Department of Physics, Govt. Degree College, Kamalpur, Dhalai, Tripura-799 285, India*
*E-mail:* `royc.partha@gmail.com`

A pure nucleonic equation of state (EoS) for beta equilibrated charge neutral neutron star (NS) matter is determined using density dependent effective NN interaction. This EoS is found to satisfy both the constraints from the observed mass-radius of neutron stars and flow data from heavy-ion collisions. Recent observations of the binary millisecond pulsar J1614-2230 by P. B. Demorest *et al.* [1] suggest that the masses lie within (1.97± 0.04) $M_o$ ($M_o$, solar mass). Most EoS involving exotic matter, such as kaon condensates or hyperons, tend to predict maximum masses well below $2.0 M_o$ and are therefore ruled out. We are able to reproduce the measured mass-radius relationship for rotating and static NS. We ensure that the star rotating not faster than the frequency limited by r-mode instability gives the maximum mass about $1.95 M_o$ with radius about 10 kilometer.

---

## 1. Introduction

A number of attempts have been made on measuring the radii and masses of neutron stars (NS) to constrain the uncertainties in the high density behavior of the equations of state. The observations on double NS [2], glitches in radio pulsars [3], thermal emission [4] from accreting NS and from millisecond X-ray pulsars lead to constraints on mass-radius (M-R) relationship of NS. Recently the pressure of neutron star matter at supranuclear density is measured by Ozel *et al.* [5] directly from observations using advanced astrophysical technique and NS atmosphere modeling. The pressure extracted from NS mass-radius data crucially constrains the extension of the EoS to high density low temperature regime for stellar matter. Certain models for the equation of state can be ruled out if those fail to reproduce the recent M-R data.

The stiffness of the high-density matter controls the maximum mass of compact stars. The analyses of M-R data on NS by Ozel *et al.* favor smaller masses lying within 1.6-1.9$M_o$ with radii 8-10 kilometers. Recent mass measurement of the binary millisecond pulsar J1614-2230 by P.B. Demorest *et al*. [1] rules out the EoS which fail to predict the masses within (1.97±0.04)$M_o$. Most of the currently proposed EoS [6-9] involving exotic matter, such as kaon condensates or hyperons are failed to produce such a massive NS. Quark matter can support a star this massive only if the quarks are strongly interacting and are therefore not 'free' quarks. To overcome this situation, Dexheimer *et al.* [10] have recently employed a hadronic SU(3) sigma-omega model including Delta-resonances and hyperons to describe the properties of neutron star matter by softer equation of state. Delta-resonances have a repulsive vector potential which works to counteract gravity in a compact star. They successfully reproduce both the measured M-R relationship and the extrapolated EoS by slightly lowering the coupling strength of the Delta resonances to the vector mesons.

In the present work, the density dependent M3Y effective interaction (DDM3Y) which provides a unified description of the elastic and the inelastic scattering, various radioactivities and nuclear matter properties, is employed to obtain nucleonic EoS of the beta equilibrated NS matter. A systematic study of the static as well as rotating NS is presented in view of the recent observations of the massive compact stars. We shall see later in the text that the present EoS unlike other EoS [6-9] can successfully reproduce the recently observed M-R data [1,5].

## 2. Isospin asymmetric nuclear matter EoS

The nuclear matter EoS is calculated [11] using the isoscalar and the isovector components of M3Y interaction along with density dependence. The density dependence of the effective interaction, DDM3Y, is completely determined from nuclear matter calculations. The equilibrium density of the nuclear matter is determined by minimizing the energy per nucleon ($\epsilon$ =E/A). The energy variation of the zero range potential is treated accurately by allowing it to vary freely with the kinetic energy part. In a Fermi gas model of interacting neutrons and protons, the $\epsilon$ for isospin asymmetric nuclear matter is the sum of the kinetic part and nuclear potential part calculated using the volume integral of the isoscalar and isovector components of M3Y interaction. The details of the present methodology may be obtained in Ref. [11].

We numerically determine the EoS for the beta equilibrated charge neutral neutron star matter in which the beta equilibrated proton fraction are determined from the knowledge of the symmetry energy. Then this EoS is used to solve the Einstein's field equations by Green's function technique to explore the various properties of rotating and static NS. The metric used in the present calculation and the code used to study the rotating stellar structure were described in a recent work by Roy Chowdhury *et al.* [11]. We consider the total energy density including mass i.e. mass-energy density which is equal to the sum of $\epsilon$ and avarage nucleonic mass (m~938.919 MeV) times baryonic number density for solving Einstein's equations for stellar structure. In Fig.1, the continuous lines represent the EoS used in this work with the saturation energy per nucleon $\epsilon_0 = -15.26 \pm 0.52$ MeV. The areas enclosed by the continuous and the dashed lines correspond to the pressure (P) regions for neutron matter (NM) consistent with the experimental flow data [12] after inclusion of the pressures from asymmetry terms with weak (soft NM) and strong (stiff NM) density dependences, respectively. It is evident from the Fig.1, that the pressure density relationship is consistent with the experimental flow data [12] and thus, confirms the high density behavior of the present EoS.

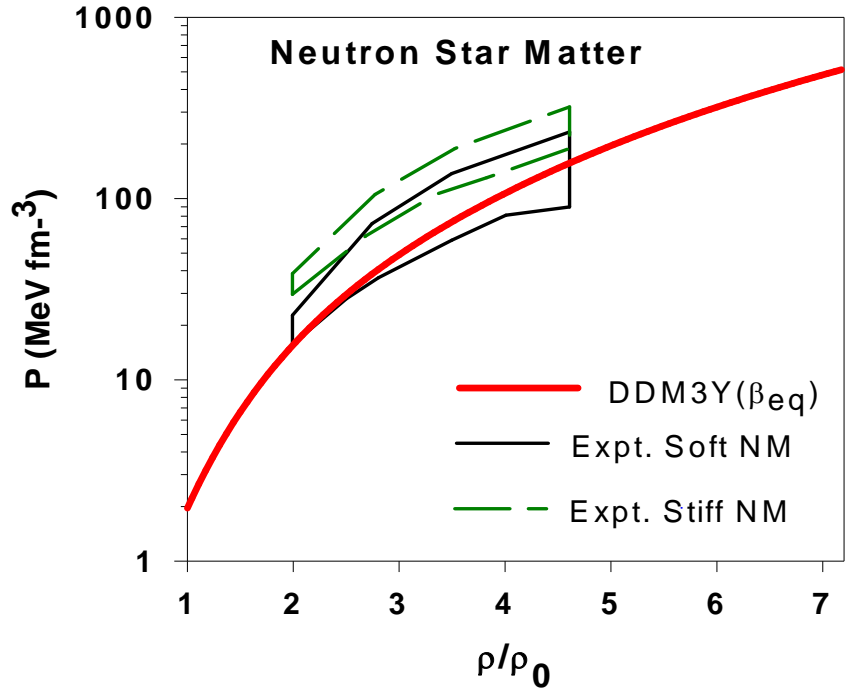

**Fig.1** *The pressure P of neutron matter as a function of density $\rho/\rho_0$.*

## 3. Results and discussion

We investigate the impact of the compression modulus and symmetry energy of nuclear matter on the maximum mass of neutron stars in view of the recent constraints from the astrophysical observations of massive neutron stars [1,13] and heavy-ion data [12,14]. The *β*-equilibrated neutron star matter using this EoS with a thin crust is able to describe highly massive compact stars (~ $2M_o$). The mass-radius relationships for the sequence of rotating stars with angular frequencies (Ω) 4.190 kHz (period T=1.5 ms) and 3.140 kHz (T=2 ms) are shown in the Fig.2. We choose those frequencies here to ensure that the time periods (T) should not be lower than the minima set by r-mode instability. Including the effect of NS rotation increases the maximum possible mass only about 2% for each EoS. Thus, our EoS provide slightly higher maximum mass for rotating NS ~ (1.93-1.95) $M_o$ with radii around 10 kilometer compared to that for static NS (~ 1.9 $M_o$) with radius about 9.6 kilometer. Due to the rotational effect, the star rotating with frequency at Keplerian limit are slightly bigger in size along the equator than the static one of same mass. The r-modes (axial fluid oscillations governed by the Coriolis force) of rapidly rotating NSs are generically unstable to the emission of gravitational waves.

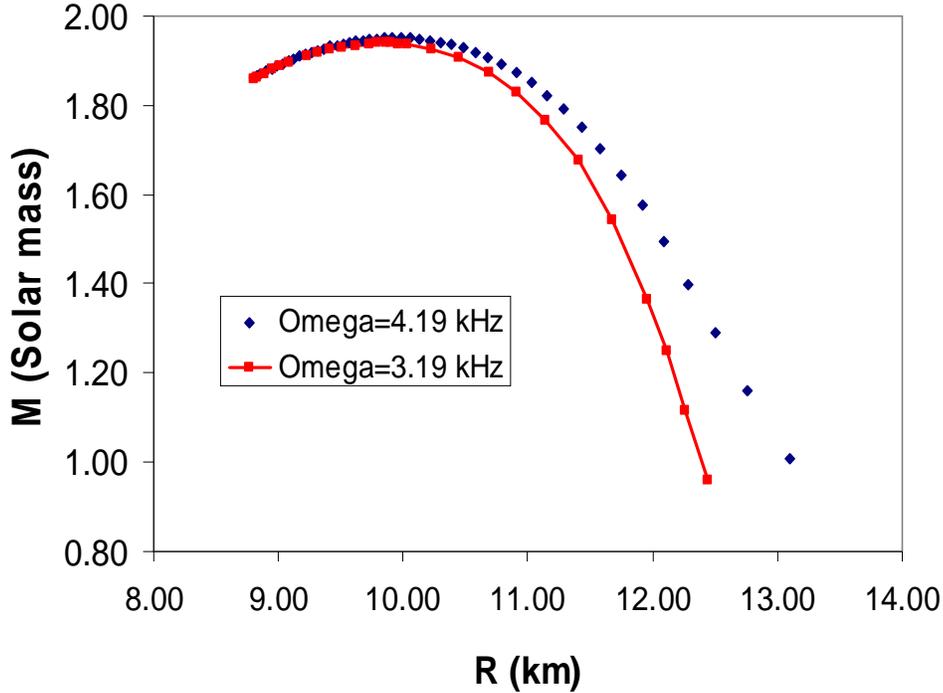

**Fig.2** *Mass-radius relationships for the sequence of stars rotating with angular frequencies (Ω) 4.19 kHz (dotted line) and 3.19 kHz (continuous line).*

The r-mode (Rossby wave) instability could slow down newly-born relativistic stars and limit their spin during accretion-induced spin-up, which would explain the absence of millisecond pulsars with rotational periods less than ~ 1.5 ms. However, the fastest-spinning neutron star PSR J17482446ad rotating with a frequency of 716 Hz (i.e. period T~1.39 ms) exceeds this limit [15]. It is also not clear whether the driving of these modes by gravitational waves can overcome viscous damping in the star. The possible importance of rapidly rotating NSs near the r-mode oscillation limit is as the gravitational wave sources for detectors like LIGO. This helps to explore the modeling of neutron star structure and gravitational wave instabilities.

**4. Summary**

We have applied our nucleonic EoS with a thin crust to solve the Einstein's field equations to determine the mass-radius relationship of neutron stars. We ensure that rotating stars used in this work are not suffered by the r-mode instability. We have obtained the rotating star mass around (1.93-1.95) $M_\odot$ with radii around 10 kilometers which are in excellent agreement with recent astrophysical observations.